\documentclass[iop,numberedappendix]{emulateapj}
\usepackage{amssymb,amsmath,graphicx,natbib,color}
\usepackage[breaklinks,colorlinks,citecolor=blue]{hyperref}

\newcommand{\be}{\begin{equation}}
\newcommand{\ee}{\end{equation}}

\newcommand{\Rph}{\ensuremath{R_\star}}

\newcommand{\Pisyn}{\ensuremath{\Pi_\mathrm{syn}}}
\newcommand{\Esyn}{\ensuremath{E_\mathrm{syn}}}

\newcommand{\me}{\ensuremath{m_\mathrm{e}}}

\newcommand{\jp}{\ensuremath{j^\prime_{\nu^\prime}}}
\newcommand{\Pp}{\ensuremath{P_{\nu^\prime}}}
\newcommand{\eB}{\ensuremath{\epsilon_\mathrm{B}}}
\newcommand{\erad}{\ensuremath{\epsilon_\mathrm{rad}}}
\newcommand{\eph}{\ensuremath{\epsilon_\star}}
\newcommand{\Ld}{\ensuremath{L_d}}
\newcommand{\D}{\ensuremath{\mathrm{d}}}
\newcommand{\nucyc}{\ensuremath{\nu^\prime_B}}
\newcommand{\numax}{\ensuremath{\nu_\mathrm{max}}}
\newcommand{\nup}{\ensuremath{\nu^\prime}}
\newcommand{\fnsc}{\ensuremath{f_\mathrm{nsc}}}

\def\Epk{E_\mathrm{pk}}
\def\Eq{Equation}
\def\nut{\nu_0}
\def\nuph{\nu_\star}
\def\Eph{E_\star}
\def\nupk{\nu_{\rm pk}}
\def\nus{\nu_\mathrm{syn}}
\def\Es{E_\mathrm{syn}}

\usepackage{ulem}

\newcommand{\KN}{\ensuremath{\xi_\mathrm{KN}}}

\newbox\grsign \setbox\grsign=\hbox{$>$} \newdimen\grdimen \grdimen=\ht\grsign
\newbox\simlessbox \newbox\simgreatbox \newbox\simpropbox
\setbox\simgreatbox=\hbox{\raise.5ex\hbox{$>$}\llap
     {\lower.5ex\hbox{$\sim$}}}\ht1=\grdimen\dp1=0pt
\setbox\simlessbox=\hbox{\raise.5ex\hbox{$<$}\llap
     {\lower.5ex\hbox{$\sim$}}}\ht2=\grdimen\dp2=0pt
\setbox\simpropbox=\hbox{\raise.5ex\hbox{$\propto$}\llap
     {\lower.5ex\hbox{$\sim$}}}\ht2=\grdimen\dp2=0pt
\def\simgt{\mathrel{\copy\simgreatbox}}

\shorttitle{Polarized photospheric emission}
\shortauthors{Lundman, Vurm \& Beloborodov}

\begin{document}

\title{Polarization of gamma-ray bursts in the dissipative photosphere model}
\author{Christoffer Lundman\altaffilmark{1,2,3}, Indrek Vurm\altaffilmark{1,4} and Andrei M. Beloborodov\altaffilmark{1}}
\affil{$^1$Physics Department and Columbia Astrophysics Laboratory, Columbia University, 538 West 120th Street, New York, \\
NY 10027, USA; clundman@particle.kth.se \\
$^2$Department of Physics, KTH Royal Institute of Technology, AlbaNova, SE-106 91 Stockholm, Sweden \\
$^3$The Oskar Klein Centre for Cosmoparticle Physics, AlbaNova, SE-106 91 Stockholm, Sweden \\
$^4$Tartu Observatory, T\~{o}ravere 61602, Tartumaa, Estonia}

\label{firstpage}







\begin{abstract}
The MeV spectral peak of gamma-ray bursts (GRBs) is best explained as photospheric emission from a dissipative relativistic jet. The observed non-blackbody spectrum shows that sub-photospheric dissipation involves both thermal plasma heating and injection of nonthermal particles, which quickly cool through inverse Compton scattering and emission of synchrotron radiation. Synchrotron photons emitted around and above the photosphere are predicted to dominate the low-energy part of the GRB spectrum, starting from roughly a decade in energy below the MeV peak. We show that this leads to a unique polarization signature: a rise in GRB polarization toward lower energies. We compute the polarization degree of GRB radiation as a function of photon energy for a generic jet model, and show the predictions for GRBs 990123, 090902B and 110721A. The expected polarization is significant in the X-ray band, in particular for bursts similar to GRB 090902B. Radiation in the MeV peak (and at higher energies) is unpolarized as long as the jet is approximately uniform on angular scales $\delta\theta \gtrsim \Gamma^{-1}$ where $\Gamma$ is the bulk Lorentz factor of the jet.
\end{abstract}




\section{Introduction}
\label{sect:introduction}

Polarization properties of gamma-ray bursts (GRBs) are poorly known, and future measurements are expected to provide important tests for the burst emission mechanism. Current polarization measurements\footnote{Specifically, GRB~041219A \citep{KalEtAl:2007, McGEtAl:2007, GotEtAl:2009}, GRB~061122 \citep{McGEtAl:2009, GotEtAl:2013}, GRB~100826A \citep{YonEtAl:2011}, GRB~110301A, GRB~110721A \citep{YonEtAl:2012}, GRB~140206A \citep{GotEtAl:2014}.} suffer from low photon statistics, although claims of detection of linear polarization degrees of a few tens of percent have been made \citep{GotEtAl:2009, YonEtAl:2011, YonEtAl:2012, GotEtAl:2013, GotEtAl:2014}.

Linear polarization is often viewed as a signature of synchrotron emission (e.g. \citealt{GotEtAl:2009, GotEtAl:2013, YonEtAl:2012}). On the other hand, several characteristics of the GRB spectrum suggest that the observed MeV peak is not synchrotron radiation. First, the spectral indices below the main MeV peak are typically harder than allowed by synchrotron emission from fast cooling electrons, and roughly half of the GRB population also violate the limit set by slow cooling electrons (e.g. \citealt{PreEtAl:1998, KanEtAl:2006, GolEtAl:2012, BurEtAl:2014}). Second, the observed distribution of the peak energy $\Epk$ is roughly log-normal and only about one order of magnitude wide (e.g. \citealt{GolEtAl:2012}). There is no \textit{a priori} reason for synchrotron radiation to produce such a narrow distribution; instead it would be expected to show a broad distribution of $\Epk$ due to its sensitivity to several parameters --- the bulk Lorentz factor of the jet, $\Gamma$, the Lorentz factor of accelerated electrons, $\gamma$, and the magnetic field strength $B$ ($\Epk \propto \Gamma \gamma^2 B$). Third, the sharpness of the observed spectral peak is inconsistent with synchrotron radiation \citep{AxeBor:2015, YuEtAl:2015, VurBel:2015}.

An alternative model has been developed, where the observed radiation is mainly produced at the early, opaque stage of jet expansion and released at its photosphere. Radiative transfer simulations demonstrate that photospheric radiation escapes with a nonthermal spectrum similar to the observed Band function \citep{BanEtAl:1993} shape (e.g. \citealt{PeeMesRee:2006, Bel:2010, VurEtAl:2011, Gia:2012}). This model naturally explains both the observed range of $\Epk$ \citep{Bel:2013} as well as the spectral shape \citep{VurBel:2015}. The hardness of the low-energy spectral index is then limited only by the Rayleigh-Jeans slope of the Planck function, consistent with all observations to date, but is expected to be typically much softer if the jet is at least moderately magnetized \citep{VurEtAl:2011} or if the jet has structure on small angular scales \citep{LunPeeRyd:2013}.

The photospheric model predicts that the MeV peak of the GRB spectrum is mainly shaped by Compton scattering of photons produced below the photosphere. The scattered radiation is intrinsically polarized (Beloborodov 2011), however the polarization of radiation received by a distant observer averages out to zero, as long as the observed outflow can be approximated as spherically symmetric. Thus, the polarization of scattered radiation can only be detected if the symmetry is broken within the patch of the jet visible to the observer, which has an angular size of $\delta\theta\sim\Gamma^{-1}$.

\citet{LunPeeRyd:2014} and \citet{ItoEtAl:2014} found that polarization degrees of up to $\Pi \sim 40 \%$ can be observed if the jet has significant structure on scales $\sim\delta\theta$, in particular if the jet is strongly beamed and its edge falls into the observed patch. Such an orientation is likely if the jet opening angle is not much larger than $\Gamma^{-1}$, so that most observers see the jet edge.

The Lorentz factors of GRB jets exceed $10^2$, and it is unclear if they can be collimated within angles  $\sim \Gamma^{-1}$. In this work we consider photospheric emission from jets without significant variations on angular scales $\delta\theta\sim\Gamma^{-1}$, which can emit polarized radiation only through the synchrotron mechanism. The magnetic field is assumed to be advected from the central engine by the expanding jet and ordered over the visible patch of a transverse angular scale $\delta\theta\sim\Gamma^{-1}$. Then the direction of the magnetic field determines the polarization plane.

\citet{VurBel:2015} recently reconstructed the radiative transfer and subphotospheric dissipation history for several GRBs by fitting simulated spectra to observed spectra. They showed that the jets are heated over a wide range in radius, typically encompassing the jet photosphere. Here we consider the same type of modeling, but we are now interested primarily in the polarization properties of the observed emission. Several versions of the dissipation mechanism have been discussed \citep{Tho:1994, EicLev:2000, DreSpr:2002, ReeMes:2005, Bel:2010, Lev:2012}. Our calculations, however, will not be specific to a particular dissipation model. We only assume that a fraction of the dissipated energy is channeled into relativistic electrons and/or positrons, as indicated by observed GRB spectra. Such nonthermal particles are expected from nuclear collisional dissipation \citep{Bel:2010} or (sub)photospheric internal shocks with significant collisionless sub-shocks \citep{Bel:2016}. Dissipation of magnetic energy through reconnection can also produce high-energy electrons, however its energy budget may be insufficient in the moderately magnetized jets that are preferred by the radiative transfer models of GRB spectra \citep{VurBel:2015}.

The injected relativistic particles emit part of their energy as polarized synchrotron emission. Photons emitted deep below the photosphere, where the scattering optical depth is much larger than unity, will necessarily scatter several times before escaping the outflow and reaching the observer. The original polarization set by the magnetic field is lost in essentially a single scattering, and therefore synchrotron photons produced deep below the photosphere will only contribute to the unpolarized part of the observed radiation. A significant fraction of synchrotron photons produced around and above the photosphere will escape without scattering and preserve their polarization. The relative contribution of these photons to the overall spectrum is a sensitive function of photon energy, as will be demonstrated below.

The goal of this paper is to make quantitative predictions for the expected GRB polarization using recent radiative transfer simulations that reconstruct the contribution of synchrotron emission to GRB spectra. We also explore which of recent bright GRBs would be most promising for the detection of polarization. The paper is organized as follows. We introduce a generic model for the energy dissipation and estimate the qualitative behaviour of the energy dependence of the observed polarization degree in Section~\ref{sect:estimates}, showing that the keV emission can be strongly polarized if the dissipation extends significantly beyond the photosphere. In Section~\ref{sect:numerical} we perform detailed numerical calculations of the polarization degree as a function of energy for dissipation parameters obtained from spectral fits of three specific GRBs (990123, 090902B and 110721A), which show that the keV emission of both GRB~990123 and GRB~090902B is expected to have been strongly polarized. Finally, we discuss our results in Section~\ref{sect:discussion}.


\section{Frequency dependence of the polarization degree}
\label{sect:estimates}

At any photon energy $E=h\nu$, the observed GRB spectrum $L_\mathrm{obs}(E)$ is the sum of two contributions: photons that escaped the jet after their last Compton scattering, $L_{\rm sc}(E)$, and photons escaping directly after their emission by the synchrotron mechanism, with no scattering, $L_{\rm nsc}(E)$. It is convenient to define the unscattered fraction,

\be
\fnsc(E)=\frac{L_{\rm nsc}(E)}{L_\mathrm{obs}(E)}.
\label{eq:f}
\ee

\noindent Then the observed polarization degree is given by

\be
\Pi(E) = \fnsc(E)\Pisyn,
\ee

\noindent where $\Pisyn$ is the polarization degree of pure unscattered synchrotron emission. Synchrotron emission from relativistic electrons in a uniform magnetic field ${\mathbf B}$ is linearly polarized in the plane perpendicular to ${\mathbf B}$. The polarization degree for an isotropic electron distribution is $\Pisyn=(p+1)/(p+7/3)$, where $p \equiv -\mathrm{d}\ln N/\mathrm{d}\ln E_e$ is the slope of the electron spectrum \citep{RybLig:1979}. This standard result is somewhat modified when the observed region is a spherical patch in a relativistic outflow carrying an ordered transverse magnetic field \citep{LyuParBla:2003}. The polarization degree of optically thin synchrotron emission from relativistic jets has been studied by several authors (e.g. \citealt{Gra:2003, GraKon:2003, NakPirWax:2003, LyuParBla:2003}; see \citealt{Laz:2006, TomEtAl:2009, Tom:2013} for reviews of GRB models which produce polarized prompt emission). The typical $\Pisyn$ varies around 50\%. Factors affecting the exact $\Pisyn$ have been studied in the previous works and will not be discussed below. In our estimates and figures we will use $\Pisyn = 50 \%$ as a typical value and focus on $\fnsc(E)$ as the key factor controlling the observed polarization.

A fluid element within the GRB jet passes through distinct radiative zones as it expands: Planck, Wien, sub-photsopheric, and optically thin \citep{Bel:2013}. The fate of a synchrotron photon depends on where it is generated:

\begin{enumerate}
\item The Planck and Wien zones have a large Compton parameter $y=4(k_B T/m_e c^2) \tau \gg 1$, where $T$ is the electron temperature, $k_B$ is the Boltzmann constant, $m_e$ is the electron mass, $c$ is the speed of light and $\tau$ is the optical depth to Thomson scattering. The condition $y \gg 1$ implies saturated Comptonization --- any new emitted synchrotron photons that avoid self-absorption and induced downscattering are quickly Comptonized to the Wien peak, reaching kinetic equilibrium with the thermal electrons. The Wien peak at the end of the Wien zone (where $y$ drops to $\sim 1$) determines the spectral peak $\Epk$ of the observed GRB \citep{Bel:2013, VurBel:2015}. Using the relation $\Epk \sim 4\Gamma k_B T$, one can roughly estimate $y \sim (\tau/\Gamma)(\Epk/m_ec^2)$, which shows that the Wien zone ends at $\tau\sim 10^2$ in a typical GRB.
\item The sub-photospheric zone at $1 \lesssim \tau \lesssim 10^2$ has $y \lesssim 1$ and here Comptonization proceeds in an unsaturated regime. Most of the synchrotron photons emitted in this zone do not reach the Wien peak and form the low-energy slope of the GRB spectrum \citep{VurBel:2015}.
\item In the optically thin zone ($\tau < 1$), most of the emitted synchrotron photons will escape without scattering and preserve their initial energy {\it as well as} polarization state.
\end{enumerate}

The radial dependence of the synchotron emissivity is controlled by the nonthermal dissipation rate. It is convenient to parameterize the dissipated power per logarithmic interval in radius by

\be
\frac{\D\Ld}{\D\ln r} = \eph L \left(\frac{r}{\Rph}\right)^k,
\label{eq:dLdlnr}
\ee

\noindent where $\Rph$ is the radius of the photosphere (where $\tau = 1$), $\eph$ is a parameter describing the strength of the dissipation at the photosphere, $L$ is the total jet luminosity, and $k$ is a power law index which determines where most of the dissipation occurs. The dissipation is assumed to occur in an extended range of radii, including the photosphere. Integration of \Eq~(\ref{eq:dLdlnr}) over the dissipation region gives $\Ld$, the total luminosity given to relativistic electrons and positrons. It does not include the thermal dissipation channel (which heats the thermal plasma with a comparable or even higher rate) as our interest here is the synchrotron emission from nonthermal particles.

The jet magnetic field is assumed to have been advected from the central engine. In the absence of magnetic energy dissipation and for conical jet expansion, the magnetic luminosity $L_B$ (the isotropic equivalent of the Poynting flux) is constant with radius, which corresponds to $B\propto (r\Gamma)^{-1}$. We parameterize the strength of the magnetic field by the ratio $\eB \equiv L_{\rm B}/L$. This gives

\be
U_B = \frac{\eB L}{4 \pi r^2 \Gamma^2 c},
\ee

\noindent where $U_B \equiv B^2/8\pi$ is the magnetic field energy density.

Besides the power of nonthermal dissipation, an important parameter is the characteristic Lorentz factor of the injected high-energy particles that dominate the synchrotron emissivity. We denote this Lorentz factor by $\gamma_0$ (measured in the jet rest frame). Dissipation through nuclear collisions produces particles with the characteristic $\gamma_0 \sim m_\pi/m_e \sim 300$, where $m_\pi$ is the pion rest mass. A moderately relativistic collisionless shock gives post-shock particles with $\gamma_0 \sim (m_p/m_e Z_\pm)$, where $Z_\pm$ is the self-regulated pair loading factor \citep{Bel:2016}. Additional acceleration mechanisms may give electrons with $\gamma > \gamma_0$, however their energy budget is significantly smaller and we will neglect their emission.\footnote{The additional particles with $\gamma>\gamma_0$ could only increase the polarization, so our estimates below will be conservative.} Our estimates will be normalized to $\gamma_0 = 300$. These electrons are in the ``fast-cooling regime'', i.e. they radiate their energy on a timescale much shorter than the jet expansion timescale.

The emitted synchrotron photons have lab-frame characteristic frequencies $\nut \approx \Gamma \gamma_0^2 \nucyc$, where 

\be
\nucyc = \frac{e B }{ 2\pi m_e c} \propto r^{-1},
\ee

\noindent is the Larmor frequency, $e$ is the electron charge, and we denote comoving frequencies with a prime to distinguish them from the unprimed lab frame frequencies. A key parameter is the characteristic frequency of synchrotron photons emitted at the Thomson scattering photosphere of radius $\Rph$,

\be
\nuph\approx \Gamma\gamma_0^2\,\frac{eB(\Rph)}{2\pi m_e c}, \qquad \Eph=h\nuph.
\ee

The bulk of photons with $\nu > \nuph$ are emitted below the photosphere (since $\nut \propto B \propto r^{-1}$), and will be scattered before escaping the jet. However, the bulk of synchrotron photons with $\nu < \nuph$ are emitted above the photosphere and will not be scattered, preserving their polarization properties.

The standard expression for the photospheric radius is given by

\be
\Rph \approx \frac{L\sigma_\mathrm{T} Z_\pm}{4 \pi m_p c^3 \Gamma^3},
\ee

\noindent where we took into account the $e^\pm$ enrichment of dissipative jets by the factor $Z_\pm$. The photosphere is quite fuzzy, as the locations of last scattering are broadly distributed around $\Rph$: 2/3 of photons propagating from large optical depths are last scattered between $0.3\Rph$ and $3\Rph$, and 1/3 --- outside this interval (\citealt{Bel:2011}; see also \citealt{Pee:2008}). The characteristic lab frame energy of a synchrotron photon emitted at $\Rph$ is

\begin{eqnarray}
\nonumber
\Eph \approx & 54  &
\left(\frac{Z_\pm}{10}\right)^{-1}
\left(\frac{\gamma_0}{300}\right)^2 \left(\frac{L}{10^{53} \, \mathrm{erg\, s^{-1}}}\right)^{-1/2} \\  & \times &
\left(\frac{\eB}{2 \times 10^{-2}}\right)^{1/2} \left(\frac{\Gamma}{500}\right)^3 \, \mathrm{keV},
\label{eq:Eph}
\end{eqnarray}

\noindent where we have used typical values representative of bright GRBs.

As long as the dissipation profile is not too steep (i.e. $k < -1/2$), the synchrotron emission at frequency $\nu$ (after integration  over all radii where dissipation occurs) peaks at the radius where the characteristic frequency $\nut$ equals $\nu$. Simplest estimates for the expected polarization can be made assuming that all weakly Comptonized synchrotron radiation at frequency $\nu$ comes from the radius where $\nut = \nu$. This approximation is reasonable for synchrotron photons emitted at $\tau \lesssim 10$. Then the radial distribution of the synchrotron spectral luminosity is given by

\be
\frac{\D L_{\nu}^{\mathrm{syn}}}{\D\ln r} \approx \frac{\eB}{\eB + \KN\erad} \frac{\D\Ld}{\D\ln r} \delta(\nu-\Gamma\gamma_0^2 \nucyc),
\label{eq:dLsdr}
\ee

\noindent where $\delta(...)$ is the delta-function, and the prefactor takes into account that only a fraction of energy given to the nonthermal particles is converted to synchrotron radiation --- the rest converts to inverse Compton (IC) radiation. This fraction is given by $\eB/(\eB + \KN\erad)$, where $\erad L$ is the part of the jet power carried by radiation, and $\KN$ is a factor which takes into account the fact that the IC cooling can be reduced due to the Klein-Nishina reduction of the Compton cross-section. As a first approximation, $\KN \approx (1 + 4 \gamma h\nu/\me c^2)^{-3/2}$, where $\gamma$ and $h\nu$ are characteristic values for the electron Lorentz factor and photon energy, respectively \citep{ModEtAl:2005}. In general, the Klein-Nishina cooling suppression is quite strong for nonthermally heated GRB jets; if the bulk Lorentz factor is $\Gamma \gtrsim 300$, the electron Lorentz factor is $\gamma \sim 300$, and the typical observed photon energy is $\sim 1$ MeV, then $\gamma h\nu/\me c^2 \gtrsim 1$, and $\KN \lesssim 10^{-1}$. Synchrotron cooling can therefore compete with IC cooling already at fairly modest values of $\eB$.

Integrating \Eq~(\ref{eq:dLsdr}) over radius, we obtain the synchrotron spectrum

\be
\nu L_{\nu}^{\mathrm{syn}} \approx \frac{\eB \eph L}{\eB + \KN\erad} \left(\frac{\nu}{\nu_*}\right)^{-k}.
\label{eq:Lsyn}
\ee

\noindent The delta-function approximation is accurate only if the resulting spectrum in \Eq~(\ref{eq:Lsyn}) is softer than the synchrotron spectrum emitted locally at a given radius by the fast-cooling electrons, $\nu L_{\nu}^{\mathrm{syn}}\propto \nu^{1/2}$. Thus one can see that \Eq~(\ref{eq:Lsyn}) is invalid for steep dissipation profiles with $k < -1/2$. In this case the production of synchrotron photons peaks deep below the photosphere at all frequencies, leading to their scattering and suppression of polarization.

The synchrotron emission should be compared with the total GRB emission, which is shaped by both Comptonized photons advected from larger optical depths and locally produced synchrotron photons. The GRB spectrum predicted by radiative transfer simulations shows a transition at low energies from the Comptonized spectrum to the ``soft excess'' dominated by weakly Comptonized synchrotron radiation. A simple (and crude) estimate for the Comptonized spectrum is a power law with a photon index $\alpha$,

\be
\nu L_{\nu} \approx \erad L \left(\frac{\nu}{\nupk}\right)^{\alpha + 2},
\qquad \nu<\nupk=\frac{\Epk}{h}.
\label{eq:LMeV}
\ee

Comparison of Equations~(\ref{eq:Lsyn}) and (\ref{eq:LMeV}) gives an estimate for the frequency $\nus$ below which the observed emission is dominated by synchrotron emission weakly affected by Comptonization. Equating (\ref{eq:Lsyn}) and (\ref{eq:LMeV}) one finds

\be
\left(\frac{\nus}{\nupk}\right)^{\alpha + k + 2} 
\approx \left(\frac{\nuph}{\nupk}\right)^k \frac{\eph \, \eB}{\erad(\eB + \KN\erad)}.
\ee

For example, GRB~090902B has $\Epk \approx 2$~MeV and $\alpha \approx -1/2$; the radiative transfer modeling performed by \citet{VurBel:2015} yielded $k \approx -0.25$, $\erad \approx 0.5$, $\eph \approx 0.4$ and $\eB \approx 2 \times 10^{-2}$. Furthermore, the peak energy was high enough for Klein-Nishina suppression of the IC cooling to be significant, with $\KN \approx 7 \times 10^{-2}$. This gives $\nus/\nupk \approx 5 \times 10^{-2}$.

If $\nus < \nuph$, the synchrotron-dominated part of the spectrum is mainly produced in the optically thin region. Generally, significant observed polarization is expected at frequencies $\nu \lesssim \min(\nus, \, \nuph)$. The condition $\nu < \nuph$ implies that the unscattered fraction of synchrotron radiation $\fnsc^{\rm syn}(\nu)$ is significant and the condition $\nu < \nus$ implies that the spectrum is synchrotron-dominated, so $\fnsc(\nu)$ defined in \Eq~(\ref{eq:f}) is approximately equal to $\fnsc^{\rm syn}(\nu)$. For the parameters of GRB~090902B, both conditions are satisfied for photon of energies $E \lesssim 200 \, \mathrm{keV}$.

The unscattered fraction of synchrotron emission $\fnsc^{\rm syn}$ is a decreasing function of $\nu/\nuph$, and it is useful to calculate this function using a more detailed synchrotron spectrum of the fast-cooling electrons and the accurate probability of photon escape from a given optical depth $\tau$ without scattering. The calculation is described in Appendices~A and B, and the result is (for $k > -1/2$)
\be
\fnsc^{\rm syn}(\nu) \approx \left(k+\frac{1}{2}\right) \left(\frac{\nu}{\nuph}\right)^{k+1/2} 
\Gamma\left[-\left(k+\frac{1}{2}\right), \frac{\nu}{\nuph}\right],
\label{eq:fsyn}
\ee
where $\Gamma[s, x]$ is the upper incomplete $\Gamma$-function (not to be confused with the bulk Lorentz factor). For $k \sim 0$, roughly a tenth of the synchrotron photons observed at $\nu \approx \nu_*$ have avoided scattering, and so the polarization degree at this frequency is modest. For a much lower frequency $\nu=10^{-2}\nuph$, \Eq~(\ref{eq:fsyn}) gives $\fnsc^{\rm syn}\approx 0.8$. The polarization degree at such frequencies is almost equal to that of optically thin synchrotron emission.

The unscattered fraction $\fnsc^{\rm syn}$ increases with decreasing $\nu$ because the lower frequency emission is produced at smaller optical depths $\tau$ --- the typical synchrotron frequency $\nut \propto B \propto r^{-1} \propto \tau$. If dissipation ends at radius $R_\mathrm{end}$, the lowest characteristic frequency $\nut$ is $\nu_\mathrm{end} = \nu_* \Rph/R_\mathrm{end}$. The unscattered fraction will then be largest at $\nu \lesssim \nu_\mathrm{end}$. For instance, in the model for GRB~090902B, if dissipation occurs up to $R_\mathrm{end} \approx 10^2 \Rph$, then the corresponding lab frame energy is $E_\mathrm{end} = E_* \Rph/R_\mathrm{end} \approx 5 \, \mathrm{keV}$.

For even lower photon energies synchrotron self-absorption may become important. The opacity due to synchrotron self-absorption, as a function of radius and comoving frequency, is computed in Appendix~\ref{sect:C}. At $r = R_\mathrm{end}$, and $\nu = \nu_\mathrm{end} \approx \Gamma \nu_\mathrm{end}^\prime$, the opacity is given by \Eq~(\ref{eq:tau_nu}),

\be
\tau_\mathrm{\nu^\prime} \approx \frac{\eB}{\eB + \KN\erad} \frac{\Gamma \eph L}{(4\pi)^2 \Rph^2 \nu_*^3 \gamma_0 m_e \tau_\mathrm{end}^{k+1}},
\ee

\noindent where $\tau_\mathrm{end} = \Rph/R_\mathrm{end}$ is the Thomson scattering optical depth at the outer dissipation radius, and we have assumed that any pair-loading of the jet has not significantly affected the $\tau \propto r^{-1}$ scaling. For the above considered values, we find $\tau_\mathrm{\nu^\prime}(R_\mathrm{end}) \approx 3 \times 10^{-5}$, so that absorption does not affect the emission much. However, since $\tau_\mathrm{\nu^\prime} \propto \nu^{-3}$ (\Eq~\ref{eq:tau_nu}), self-absorption will become significant at lower frequencies. On the other hand, at observed energies of $\lesssim 1 \, \mathrm{keV}$, Galactic absorption is also significant.


\section{Numerical models for three bright GRBs}
\label{sect:numerical}

The polarization degree $\Pi(E)$ may be predicted for a detected GRB using its observed spectrum and its numerical model obtained from radiative transfer simulations. The transfer simulations (i) allow one to approximately reconstruct the jet magnetization and the radial distribution of the nonthermal dissipation rate, which control the synchrotron emissivity, (ii) give the photospheric radius $\Rph$, and (iii) show the relative contribution of unscattered synchrotron emission $\fnsc(E)$ to the total observed spectrum (whose peak is dominated by the Comptonized radiation advected from large optical depths).

A significant role is played by $e^\pm$ pair creation, as it increases $\Rph$ and reduces $\Eph$. In addition, pair creation affects the synchrotron spectrum produced by the high-energy particles. The standard synchrotron spectrum from fast-cooling particles injected with a fixed $\gamma_0$ is affected by both the competition between synchrotron and IC cooling (with important Klein-Nishina corrections) and the reprocessing of IC radiation into secondary $e^\pm$ pairs created in the $e^\pm$ cascade. The cascade can only be suppressed by synchrotron cooling when $\eB \gtrsim \KN\erad$. High magnetization therefore increases the polarization degree due to two separate effects: (i) more synchrotron emission is produced, and (ii) the cascade is weaker, so the pair loading and $\Rph$ are reduced, leading to a higher $\Eph$ and opening a broader spectral window $E < \Eph$ for potentially strong polarization. The transfer problem is in general highly non-linear and requires simulations which self-consistently include $e^\pm$ creation in photon-photon collisions.

\citet{VurBel:2015} reconstructed the observed spectra of GRBs 990123, 090902B and 130427A with theoretical spectra obtained by detailed simulations of radiative transfer in a dissipative jet. They used a numerical code that solves the kinetic equations for the electron and photon distribution functions and follows their self-consistent evolution in the expanding jet. The initial version of the kinetic code was designed for static sources \citep{VurPou:2009} and then developed to simulate relativistic jets \citep{VurEtAl:2011} by solving the radiative transfer equation \citep{Bel:2011}. The most recent version of the code \citep{VurBel:2015} follows the jet evolution from very large optical depths $\tau \simgt 10^3$ and calculates all relevant radiative processes, including synchrotron self-absorption, induced down-scattering, $e^\pm$ creation in photon-photon collisions, double Compton scattering, and bremsstrahlung. The simulations also follow the jet acceleration by radiation pressure.

Here we study two of the previously simulated GRBs --- GRB~990123 and GRB~090902B, using the best fit parameters from \citet{VurBel:2015}. For each burst, we identify the unscattered synchrotron component in the emitted spectrum $L_{\rm nsc}(E)$ and then find the polarization degree according
to \Eq~(\ref{eq:f}), where $L_\mathrm{obs}(E)$ is the total spectrum predicted by the transfer simulations. The unscattered synchrotron luminosity is calculated numerically using the known radial dependence of the synchrotron emissivity from the transfer simulations, and the probability for photon escape without scattering (see Appendicies~A and B).

Additionally, we include GRB~110721A in our sample, a GRB with a claimed prompt emission polarization detection. \citet{YonEtAl:2012} reported a time-integrated polarization degree of $\Pi = 84^{+16}_{-28} \%$ with a $3.3\sigma$ confidence level. We first find a radiative transfer model that reproduces the observed spectrum (we used the data from time bin 4, as presented in \citet{AxeEtAl:2012}, and assumed a cosmological redshift of $z = 2$). Then we use this model to obtain $L_{\rm nsc}(E)$ in the same way as for GRB~990123 and GRB~090902B.

The fitted dissipation parameters of each GRB are listed in Table~\ref{tab:1}. The table only shows {\it nonthermal} dissipation parameters (which is of main interest for us here) and omits the thermal heating rate, which was also part of the simulation. For details of the radiative transfer simulations, see \citet{VurBel:2015}.

\begin{table}
\centering
\caption{Fitted nonthermal dissipation parameters ($\eB$, $\eph$, $\tau_\mathrm{end}$ and $k$) and derived characteristic energies ($\Epk$ and $\Eph$)} for GRB~990123, GRB~090902B and GRB~110721A.
\begin{tabular}{l*{3}{r}}
Parameter                  &           GRB~990123 &          GRB~090902B &          GRB~110721A \\
\hline\\[-10pt]
$\eB$                      & $1.8 \times 10^{-2}$ & $1.1 \times 10^{-2}$ & $1.0 \times 10^{-3}$ \\
$\eph$                     & $6.8 \times 10^{-3}$ & $1.5 \times 10^{-2}$ & $2.3 \times 10^{-2}$ \\
$\tau_\mathrm{end}$        & $1.2 \times 10^{-2}$ & $3.0 \times 10^{-2}$ & $4.0 \times 10^{-1}$ \\
$k$                        &              $-0.19$ &              $-0.25$ &             $-0.013$ \\
$\Epk$ (MeV)               &                  1.4 &                  2.6 &                  1.0 \\
$\Eph$ (keV)               &                   69 &                  140 &                 0.88 \\
\end{tabular}
\label{tab:1}
\end{table}

\begin{figure}
\centering
\includegraphics[width=\linewidth]{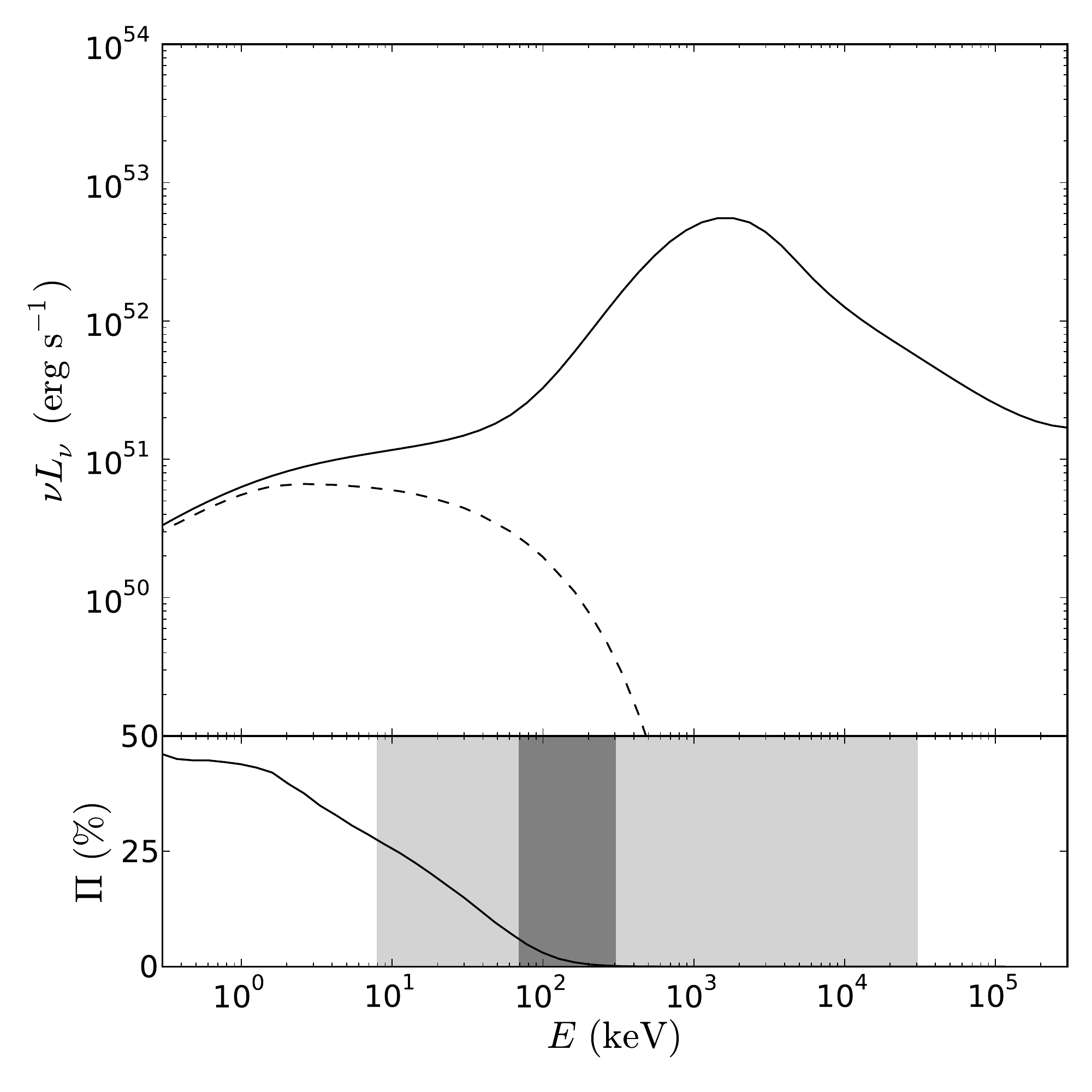}
\caption{Top panel: the simulated GRB spectrum for GRB~990123 (solid line) and the spectrum of unscattered synchrotron emission (dashed line).
Bottom panel: the polarization degree (i.e. the ratio of the above spectra times the assumed synchrotron polarization degree of $50 \%$) as a function of energy. The light and dark shaded regions correspond to the \textit{Fermi} GBM (NaI + BGO detectors, 8~keV to 30~MeV) and \textit{GAP} (70 to 300~keV) energy ranges respectively.}
\label{fig:990123}
\end{figure}

\begin{figure}
\centering
\includegraphics[width=\linewidth]{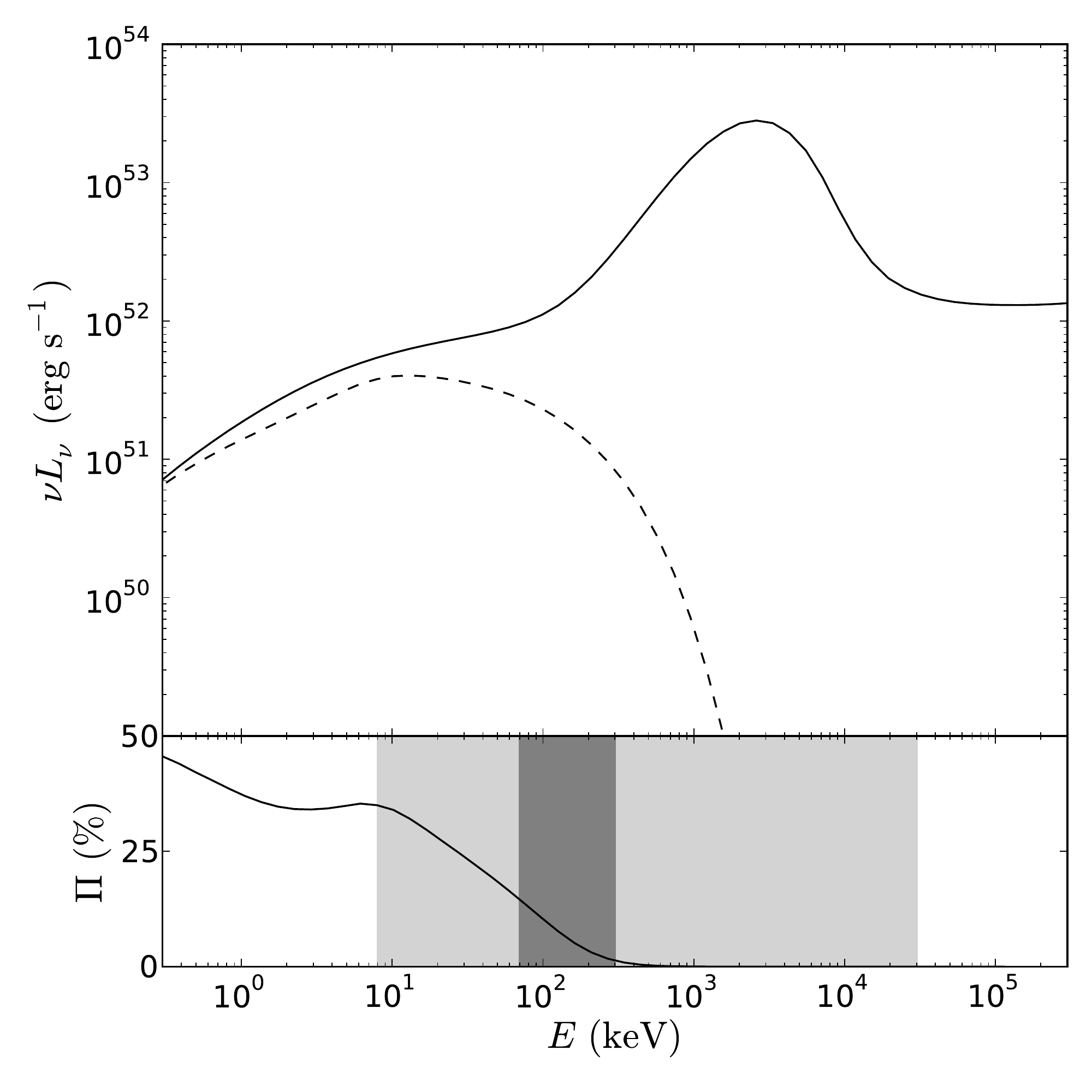}
\caption{Same as Figure~\ref{fig:990123}, but for GRB~090902B.}
\label{fig:090902B}
\end{figure}

\begin{figure}
\centering
\includegraphics[width=\linewidth]{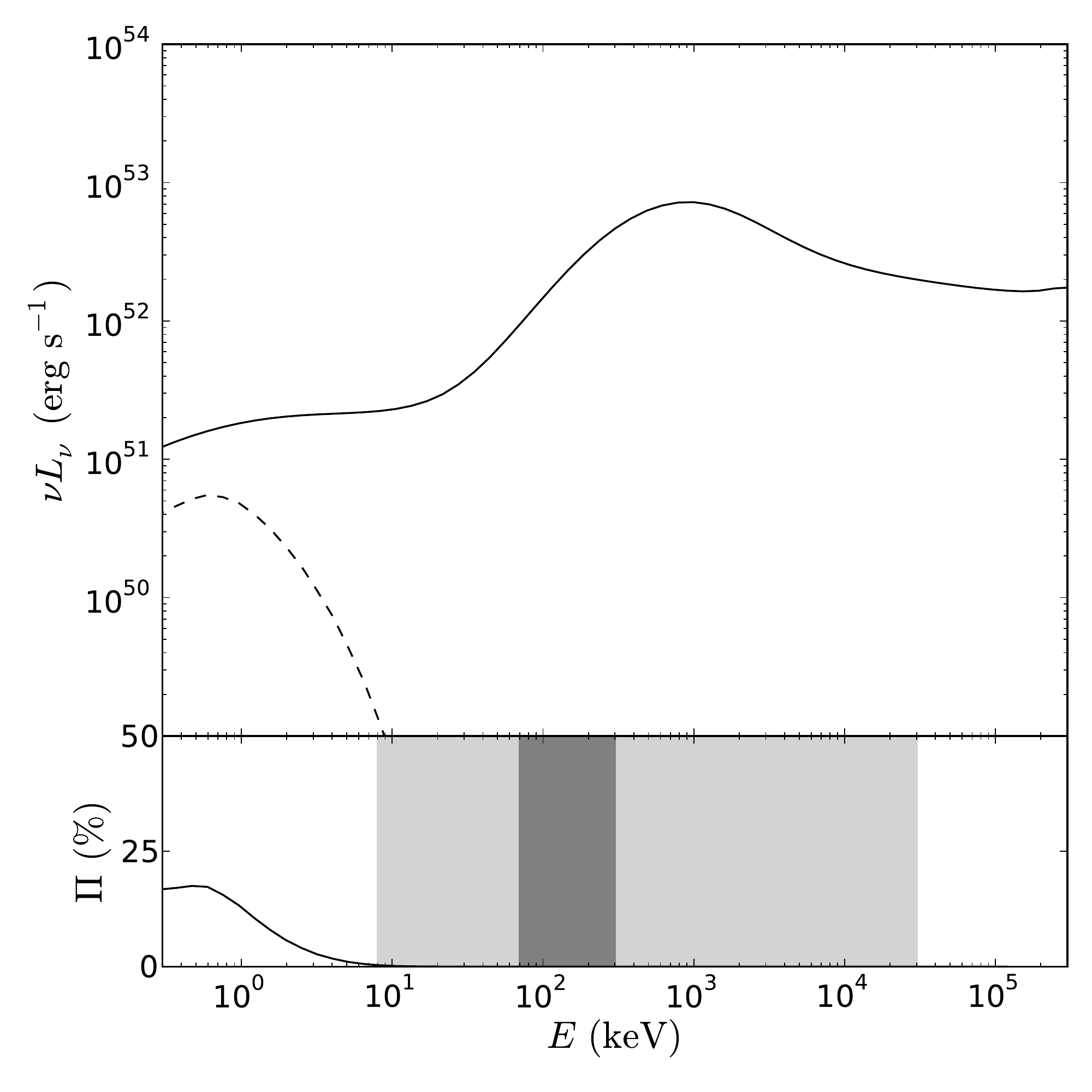}
\caption{Same as Figure~\ref{fig:990123}, but for GRB~110721A.}
\label{fig:110721A}
\end{figure}

The numerically integrated unscattered synchrotron spectra, as well as the total GRB spectra from the radiative transfer simulations, are shown in Figures~\ref{fig:990123}-\ref{fig:110721A}. These figures demonstrate the essential features discussed in Section~\ref{sect:estimates}. The MeV peak is unpolarized, as it was formed in regions of large optical depths. Similarly, the spectrum above the peak consists of Comptonized photons which are also unpolarized. Synchrotron emission dominates the spectrum only at low energies, and only a fraction of this emission has avoided scattering before escaping the jet.

In order to observe a significant polarization degree, significant nonthermal dissipation must occur near and above the photosphere. This is the case for GRB~990123 and GRB~090902B (Figures~\ref{fig:990123} and \ref{fig:090902B} respectively), which are best modeled by rather flat dissipation profiles ($k \approx -1/5$ and $k \approx -1/4$). The reconstructed properties of these two GRBs are qualitatively similar, resulting in similar spectral features. The reconstructed magnetization, $\eB \sim 10^{-2}$, is strong enough to partially suppress the pair cascade, so that the increase of $\Rph$ due to pair loading is moderate. The partial suppression of the cascade also manifests itself in weaker, less Comptonized high energy spectra. The full spectra show curvature at $E \sim \Es \sim 100 \, \mathrm{keV}$, which coincides with $\Eph$, indicating the transition to optically thin synchrotron dominated spectra. At the lowest energies ($E \lesssim 10 \, \mathrm{keV}$) the spectra curve downwards as a result of synchrotron self-absorption. The observed polarization degrees are a few tens of percent at $10 - 100$ keV.

The best spectral fit to GRB~110721A (Figure~\ref{fig:110721A}) has an almost flat dissipation profile across the photosphere ($k \approx 0$). The rather weak magnetization $\eB \approx 10^{-3}$ results in a fully developed IC $e^\pm$ cascade. The photosphere was therefore pushed further out, and the characteristic synchrotron energy at the photosphere $\Eph$ was signficantly reduced. The resulting polarization degree is significant only at $E \lesssim 3$ keV.

\section{Discussion}
\label{sect:discussion}

\subsection{General conditions for polarized photospheric emission}

The above examples illustrate the conditions leading to significant polarization of GRB emission. First, nonthermal dissipation is required close to the jet photosphere ($\eph \gtrsim 10^{-2}$), involving injection of nonthermal electrons or positrons. This is expected in GRBs and consistent with their observed spectra \citep{VurBel:2015}. Second, dissipation should not decline too quickly above the photosphere ($k \gtrsim -1/2$), so that synchrotron emission extends into regions of moderate optical depth. Third, the jet must be significantly magnetized ($\eB \gtrsim \KN\erad$) in order to generate a strong synchrotron component. The significant magnetization also weakens the pair cascade, avoiding a dramatic increase of $\Rph$ by pair loading. Under such conditions, polarization degrees of a few tens of percent at observed energies of a few tens of keV is expected.

Our calculations suggest a relation between the polarization degree $\Pi(E)$ and the observed spectral shape. The strongest polarization is expected in bursts similar to GRB~090902B, where nonthermal dissipation and the synchrotron component are strong around and above the photosphere. In all of our calculated models, the spectral peak at $\Epk\sim 1$~MeV is very weakly polarized, because its formation involves multiple scattering below the photosphere, suppressing polarization. This is consistent with the observed sharpness of the MeV peak, which rules out its synchrotron origin \citep{Bel:2013,AxeBor:2015,YuEtAl:2015,VurBel:2015}.

A detection of strong polarization of the MeV peak would indicate a significant angular structure of the jet. Then the polarized signal must be due to the geometry of the scattering process, independent of magnetic fields or energy dissipation. In particular, strong polarization across the spectral peak is expected when a significant fraction of observed radiation is emitted near the edge of the collimated jet, $\theta_j$. The characteristic solid angle occupied by this radiation is $\Delta\Omega_\mathrm{edge} \sim 2\pi[\cos(\theta_\mathrm{j} - \Gamma^{-1}) - \cos(\theta_\mathrm{j} + \Gamma^{-1})]$, where we took into account the Doppler beaming of radiation within angle $\delta\theta\sim\Gamma^{-1}$. The total solid angle occupied by radiation from all $\theta<\theta_j$ is $\Delta\Omega_\mathrm{tot} \sim 2\pi[1 - \cos(\theta_\mathrm{j} + \Gamma^{-1})]$, and the probability of observing the edge may be estimated as 

$$
P_\mathrm{edge} \approx \frac{\Delta\Omega_\mathrm{edge}}{\Delta\Omega_\mathrm{tot}} 
    \sim \frac{4}{\Gamma\theta_\mathrm{j}}, 
$$

\noindent where $\cos\theta \approx 1 - \theta^2/2$ has been used. Substituting plausible values of $\theta_\mathrm{j} \sim 0.1$ and $\Gamma \sim 400$ as an example, one finds that the edge is visible in roughly one tenth of GRBs, so every tenth burst would be strongly polarized. Note however that the actual distribution of jet opening angles is uncertain; the existing estimates inferred from so-called ``jet breaks'' in the light curves of GRB afterglows do not suggest a preferred $\theta_j$, and in many cases no jet break was detected (see e.g. \citealt{RacEtAl:2009}). Polarization studies 
of the prompt emission above 100~keV could provide a new way to constrain $\theta_j$.

Our results demonstrate that in the absence of the edge effects, a moderate polarization is still expected due to the synchrotron component in GRB emission. This polarization can be significant and has the characteristic rise toward soft energies. We note however that there is an additional factor that may hinder the detection of synchrotron polarization: its fast variations on unresolved timescales. The radial profile of the jet must be strongly variable, as evidenced by the observed light curves of GRBs, and the magnetic field ejected by the central engine may be strongly variable. For instance, it may alternate on a small radial scale $\delta r$, resembling the striped winds from pulsars, and polarization may be measured during a time interval $\Delta t \gg \delta r/c$. Such observations can only probe the time-averaged degree of polarization, which could be much smaller than the instantaneous value.

\subsection{Comparison to current polarization measurements}

To date, observations of GRB polarization were performed by \textit{INTEGRAL} and \textit{GAP} satellites. In contrast to \textit{INTEGRAL}, \textit{GAP} was specifically designed and optimized to measure the polarization properties of prompt GRB emission. Both instruments utilize the polarization dependence of the Compton scattering cross-section in order to detect a polarized signal. Specifically, photons which first scatter, and then interact with the detector again are registered and used to reconstruct the polarization degree of the incoming signal. By recording the position of both interactions within the detector, and using knowledge of the direction to the GRB, one can reconstruct the distribution of azimuthal scattering angles. The modulation of the distribution is then used to reconstruct the polarization properties of the incoming signal. The requirement of subsequent interactions significantly lowers the effective area of such polarization detectors. The photon statistics are therefore poor in general, with current measurements registering at most a few thousand double events during a GRB. Due to the poor statistics, the polarization measurements are typically time integrated over the burst duration.

The reported polarization degrees are in general large. The best fit values appear roughly uniformly distributed between the smallest value of $25 \%$ (GRB~100826A, \citealt{YonEtAl:2011}) to the largest value of $84 \%$ (GRB~110721A, \citealt{YonEtAl:2012}). The 68 percent confidence intervals are typically reported to be about $\pm 20 \%$. We note that measuring the polarization of prompt GRB emission is challenging. As mentioned above, the detector effective area is usually small. Non-trivial systematic effects occur due to the fact that the experimental setup is not axially symmetric around the line-of-sight to the GRB; if not properly accounted for, these effects can mimic the modulation curve of a polarized signal and increase the measurement uncertainty. Furthermore, the polarization degree is a positive-definite quantity. Therefore, the expected value of a measurement will always be larger than zero, also for an unpolarized signal \citep{WeiElsODe:2010}. Specifically, if $\Delta\Pi$ is the typical uncertainty of a polarization degree measurement, then one expects the measured value to be some order unity fraction of $\Delta\Pi$ also when the signal is unpolarized: $\Pi \lesssim \Delta\Pi$. A better signal-to-noise ratio would be highly desirable in future observations.

It is also difficult to produce such large polarization degrees from a theoretical perspective, especially after averaging over long time-intervals, comparable to the burst duration. A confirmation of observed values $\Pi \gtrsim 60 \%$ by future measurements with large signal-to-noise ratios would be truly spectacular and extremely constraining. Polarization degrees of $\sim 50$\% could, in principle, be compatible with optically thin synchrotron emission. However, as mentioned in Section~\ref{sect:introduction}, optically thin synchrotron emission models struggle to explain the key spectral features of GRBs.

When a realistic photospheric model is used to fit the GRB spectrum, the contribution of unscattered synchrotron emission is found to be small for most bursts. As an example, we considered GRB~110721A. \citet{YonEtAl:2012} reported a polarization degree of $\Pi = 84^{+16}_{-28}$, with a non-zero detection claimed at a confidence level of $3.3 \sigma$ in the energy range 70 - 300~keV for the same burst. In contrast, our simulations give a very weak polarization degree in this energy band (Figure~3). Our spectrum reconstruction for GRB~110721A, which includes its strong emission at $E \gg 1$~MeV, suggests a fully developed pair cascade. The cascade increases the number of pairs in the jet and pushes the photosphere to larger radii, where the magnetic field is weaker and synchrotron photons have lower energies, $\Esyn \lesssim 1$~keV. As a result, the simulations predict significant polarization only in the soft X-ray band. We note however the limitations of the available spectral fits; a detailed reconstruction of bursts with well measured broad-band spectra will be key for accurate polarization predictions.

It is also important to note that detectors using Compton scattering to measure polarization do not weigh the measured polarization degree by photon energy. Each recorded double event carries the same weight for constructing the modulation curve, from which the polarization degree and the position angle are computed. If low-energy photons dominate the observed detector counts, then the measured polarization degree is also dominated by the low-energy photons. This fact should be accounted for when integrating the predicted polarization degree within a specific detector energy range.


\acknowledgments
CL acknowledges the Swedish Research Council for financial support. AMB is supported by NSF grant AST-1412485, NASA grant NNX15AE26G,
and a grant from the Simons Foundation (\#446228, Andrei Beloborodov).


\bibliographystyle{apj}
\bibliography{refgrb}

\begin{thebibliography}{}
\expandafter\ifx\csname natexlab\endcsname\relax\def\natexlab#1{#1}\fi

\bibitem[{{Axelsson} \& {Borgonovo}(2015)}]{AxeBor:2015}
{Axelsson}, M., \& {Borgonovo}, L. 2015, \mnras, 447, 3150

\bibitem[{{Axelsson} {et~al.}(2012){Axelsson}, {Baldini}, {Barbiellini},
  {Baring}, {Bellazzini}, {Bregeon}, {Brigida}, {Bruel}, {Buehler},
  {Caliandro}, {Cameron}, {Caraveo}, {Cecchi}, {Chaves}, {Chekhtman}, {Chiang},
  {Claus}, {Conrad}, {Cutini}, {D'Ammando}, {de Palma}, {Dermer}, {Silva},
  {Drell}, {Favuzzi}, {Fegan}, {Ferrara}, {Focke}, {Fukazawa}, {Fusco},
  {Gargano}, {Gasparrini}, {Gehrels}, {Germani}, {Giglietto}, {Giroletti},
  {Godfrey}, {Guiriec}, {Hadasch}, {Hanabata}, {Hayashida}, {Hou}, {Iyyani},
  {Jackson}, {Kocevski}, {Kuss}, {Larsson}, {Larsson}, {Longo}, {Loparco},
  {Lundman}, {Mazziotta}, {McEnery}, {Mizuno}, {Monzani}, {Moretti},
  {Morselli}, {Murgia}, {Nuss}, {Nymark}, {Ohno}, {Omodei}, {Pesce-Rollins},
  {Piron}, {Pivato}, {Racusin}, {Rain{\`o}}, {Razzano}, {Razzaque}, {Reimer},
  {Roth}, {Ryde}, {Sanchez}, {Sgr{\`o}}, {Siskind}, {Spandre}, {Spinelli},
  {Stamatikos}, {Tibaldo}, {Tinivella}, {Usher}, {Vandenbroucke}, {Vasileiou},
  {Vianello}, {Vitale}, {Waite}, {Winer}, {Wood}, {Burgess}, {Bhat},
  {Bissaldi}, {Briggs}, {Connaughton}, {Fishman}, {Fitzpatrick}, {Foley},
  {Gruber}, {Kippen}, {Kouveliotou}, {Jenke}, {McBreen}, {McGlynn}, {Meegan},
  {Paciesas}, {Pelassa}, {Preece}, {Tierney}, {von Kienlin}, {Wilson-Hodge},
  {Xiong}, \& {Pe'er}}]{AxeEtAl:2012}
{Axelsson}, M., {Baldini}, L., {Barbiellini}, G., {et~al.} 2012, \apjl, 757,
  L31

\bibitem[{{Band} {et~al.}(1993){Band}, {Matteson}, {Ford}, {Schaefer},
  {Palmer}, {Teegarden}, {Cline}, {Briggs}, {Paciesas}, {Pendleton}, {Fishman},
  {Kouveliotou}, {Meegan}, {Wilson}, \& {Lestrade}}]{BanEtAl:1993}
{Band}, D., {Matteson}, J., {Ford}, L., {et~al.} 1993, \apj, 413, 281

\bibitem[{{Beloborodov}(2010)}]{Bel:2010}
{Beloborodov}, A.~M. 2010, \mnras, 407, 1033

\bibitem[{{Beloborodov}(2011)}]{Bel:2011}
---. 2011, \apj, 737, 68

\bibitem[{{Beloborodov}(2013)}]{Bel:2013}
---. 2013, \apj, 764, 157

\bibitem[{{Beloborodov}(2016)}]{Bel:2016}
---. 2016, ArXiv e-prints, arXiv:1604.02794

\bibitem[{{Burgess} {et~al.}(2014){Burgess}, {Ryde}, \& {Yu}}]{BurEtAl:2014}
{Burgess}, J.~M., {Ryde}, F., \& {Yu}, H.-F. 2014, ArXiv e-prints,
  arXiv:1410.7647

\bibitem[{{Drenkhahn} \& {Spruit}(2002)}]{DreSpr:2002}
{Drenkhahn}, G., \& {Spruit}, H.~C. 2002, \aap, 391, 1141

\bibitem[{{Eichler} \& {Levinson}(2000)}]{EicLev:2000}
{Eichler}, D., \& {Levinson}, A. 2000, \apj, 529, 146

\bibitem[{{Ghisellini} \& {Svensson}(1991)}]{GhiSve:1991}
{Ghisellini}, G., \& {Svensson}, R. 1991, \mnras, 252, 313

\bibitem[{{Giannios}(2012)}]{Gia:2012}
{Giannios}, D. 2012, \mnras, 422, 3092

\bibitem[{{Goldstein} {et~al.}(2012){Goldstein}, {Burgess}, {Preece}, {Briggs},
  {Guiriec}, {van der Horst}, {Connaughton}, {Wilson-Hodge}, {Paciesas},
  {Meegan}, {von Kienlin}, {Bhat}, {Bissaldi}, {Chaplin}, {Diehl}, {Fishman},
  {Fitzpatrick}, {Foley}, {Gibby}, {Giles}, {Greiner}, {Gruber}, {Kippen},
  {Kouveliotou}, {McBreen}, {McGlynn}, {Rau}, \& {Tierney}}]{GolEtAl:2012}
{Goldstein}, A., {Burgess}, J.~M., {Preece}, R.~D., {et~al.} 2012, \apjs, 199,
  19

\bibitem[{{G{\"o}tz} {et~al.}(2013){G{\"o}tz}, {Covino}, {Fern{\'a}ndez-Soto},
  {Laurent}, \& {Bo{\v s}njak}}]{GotEtAl:2013}
{G{\"o}tz}, D., {Covino}, S., {Fern{\'a}ndez-Soto}, A., {Laurent}, P., \&
  {Bo{\v s}njak}, {\v Z}. 2013, \mnras, 431, 3550

\bibitem[{{G{\"o}tz} {et~al.}(2014){G{\"o}tz}, {Laurent}, {Antier}, {Covino},
  {D'Avanzo}, {D'Elia}, \& {Melandri}}]{GotEtAl:2014}
{G{\"o}tz}, D., {Laurent}, P., {Antier}, S., {et~al.} 2014, \mnras, 444, 2776

\bibitem[{{G{\"o}tz} {et~al.}(2009){G{\"o}tz}, {Laurent}, {Lebrun}, {Daigne},
  \& {Bo{\v s}njak}}]{GotEtAl:2009}
{G{\"o}tz}, D., {Laurent}, P., {Lebrun}, F., {Daigne}, F., \& {Bo{\v s}njak},
  {\v Z}. 2009, \apjl, 695, L208

\bibitem[{{Granot}(2003)}]{Gra:2003}
{Granot}, J. 2003, \apjl, 596, L17

\bibitem[{{Granot} \& {K{\"o}nigl}(2003)}]{GraKon:2003}
{Granot}, J., \& {K{\"o}nigl}, A. 2003, \apjl, 594, L83

\bibitem[{{Ito} {et~al.}(2014){Ito}, {Nagataki}, {Matsumoto}, {Lee}, {Tolstov},
  {Mao}, {Dainotti}, \& {Mizuta}}]{ItoEtAl:2014}
{Ito}, H., {Nagataki}, S., {Matsumoto}, J., {et~al.} 2014, \apj, 789, 159

\bibitem[{{Kalemci} {et~al.}(2007){Kalemci}, {Boggs}, {Kouveliotou}, {Finger},
  \& {Baring}}]{KalEtAl:2007}
{Kalemci}, E., {Boggs}, S.~E., {Kouveliotou}, C., {Finger}, M., \& {Baring},
  M.~G. 2007, \apjs, 169, 75

\bibitem[{{Kaneko} {et~al.}(2006){Kaneko}, {Preece}, {Briggs}, {Paciesas},
  {Meegan}, \& {Band}}]{KanEtAl:2006}
{Kaneko}, Y., {Preece}, R.~D., {Briggs}, M.~S., {et~al.} 2006, \apjs, 166, 298

\bibitem[{{Lazzati}(2006)}]{Laz:2006}
{Lazzati}, D. 2006, New Journal of Physics, 8, 131

\bibitem[{{Levinson}(2012)}]{Lev:2012}
{Levinson}, A. 2012, \apj, 756, 174

\bibitem[{{Lundman} {et~al.}(2013){Lundman}, {Pe'er}, \&
  {Ryde}}]{LunPeeRyd:2013}
{Lundman}, C., {Pe'er}, A., \& {Ryde}, F. 2013, \mnras, 428, 2430

\bibitem[{{Lundman} {et~al.}(2014){Lundman}, {Pe'er}, \&
  {Ryde}}]{LunPeeRyd:2014}
---. 2014, \mnras, 440, 3292

\bibitem[{{Lyutikov} {et~al.}(2003){Lyutikov}, {Pariev}, \&
  {Blandford}}]{LyuParBla:2003}
{Lyutikov}, M., {Pariev}, V.~I., \& {Blandford}, R.~D. 2003, \apj, 597, 998

\bibitem[{{McGlynn} {et~al.}(2007){McGlynn}, {Clark}, {Dean}, {Hanlon},
  {McBreen}, {Willis}, {McBreen}, {Bird}, \& {Foley}}]{McGEtAl:2007}
{McGlynn}, S., {Clark}, D.~J., {Dean}, A.~J., {et~al.} 2007, \aap, 466, 895

\bibitem[{{McGlynn} {et~al.}(2009){McGlynn}, {Foley}, {McBreen}, {Hanlon},
  {McBreen}, {Clark}, {Dean}, {Martin-Carrillo}, \& {O'Connor}}]{McGEtAl:2009}
{McGlynn}, S., {Foley}, S., {McBreen}, B., {et~al.} 2009, \aap, 499, 465

\bibitem[{{Moderski} {et~al.}(2005){Moderski}, {Sikora}, {Coppi}, \&
  {Aharonian}}]{ModEtAl:2005}
{Moderski}, R., {Sikora}, M., {Coppi}, P.~S., \& {Aharonian}, F. 2005, \mnras,
  363, 954

\bibitem[{{Nakar} {et~al.}(2003){Nakar}, {Piran}, \& {Waxman}}]{NakPirWax:2003}
{Nakar}, E., {Piran}, T., \& {Waxman}, E. 2003, \jcap, 10, 5

\bibitem[{{Pe'er}(2008)}]{Pee:2008}
{Pe'er}, A. 2008, \apj, 682, 463

\bibitem[{{Pe'er} {et~al.}(2006){Pe'er}, {M{\'e}sz{\'a}ros}, \&
  {Rees}}]{PeeMesRee:2006}
{Pe'er}, A., {M{\'e}sz{\'a}ros}, P., \& {Rees}, M.~J. 2006, \apj, 642, 995

\bibitem[{{Preece} {et~al.}(1998){Preece}, {Briggs}, {Mallozzi}, {Pendleton},
  {Paciesas}, \& {Band}}]{PreEtAl:1998}
{Preece}, R.~D., {Briggs}, M.~S., {Mallozzi}, R.~S., {et~al.} 1998, \apjl, 506,
  L23

\bibitem[{{Racusin} {et~al.}(2009){Racusin}, {Liang}, {Burrows}, {Falcone},
  {Sakamoto}, {Zhang}, {Zhang}, {Evans}, \& {Osborne}}]{RacEtAl:2009}
{Racusin}, J.~L., {Liang}, E.~W., {Burrows}, D.~N., {et~al.} 2009, \apj, 698,
  43

\bibitem[{{Rees} \& {M{\'e}sz{\'a}ros}(2005)}]{ReeMes:2005}
{Rees}, M.~J., \& {M{\'e}sz{\'a}ros}, P. 2005, \apj, 628, 847

\bibitem[{{Rybicki} \& {Lightman}(1979)}]{RybLig:1979}
{Rybicki}, G.~B., \& {Lightman}, A.~P. 1979, {Radiative processes in
  astrophysics} (Wiley-VCH)

\bibitem[{{Thompson}(1994)}]{Tho:1994}
{Thompson}, C. 1994, \mnras, 270, 480

\bibitem[{{Toma}(2013)}]{Tom:2013}
{Toma}, K. 2013, ArXiv e-prints, arXiv:1308.5733

\bibitem[{{Toma} {et~al.}(2009){Toma}, {Sakamoto}, {Zhang}, {Hill},
  {McConnell}, {Bloser}, {Yamazaki}, {Ioka}, \& {Nakamura}}]{TomEtAl:2009}
{Toma}, K., {Sakamoto}, T., {Zhang}, B., {et~al.} 2009, \apj, 698, 1042

\bibitem[{{Vurm} \& {Beloborodov}(2015)}]{VurBel:2015}
{Vurm}, I., \& {Beloborodov}, A.~M. 2015, ArXiv e-prints, arXiv:1506.01107

\bibitem[{{Vurm} {et~al.}(2011){Vurm}, {Beloborodov}, \&
  {Poutanen}}]{VurEtAl:2011}
{Vurm}, I., {Beloborodov}, A.~M., \& {Poutanen}, J. 2011, \apj, 738, 77

\bibitem[{{Vurm} \& {Poutanen}(2009)}]{VurPou:2009}
{Vurm}, I., \& {Poutanen}, J. 2009, \apj, 698, 293

\bibitem[{{Weisskopf} {et~al.}(2010){Weisskopf}, {Elsner}, \&
  {O'Dell}}]{WeiElsODe:2010}
{Weisskopf}, M.~C., {Elsner}, R.~F., \& {O'Dell}, S.~L. 2010, in \procspie,
  Vol. 7732, Space Telescopes and Instrumentation 2010: Ultraviolet to Gamma
  Ray, 77320E

\bibitem[{{Yonetoku} {et~al.}(2011){Yonetoku}, {Murakami}, {Gunji}, {Mihara},
  {Toma}, {Sakashita}, {Morihara}, {Takahashi}, {Toukairin}, {Fujimoto},
  {Kodama}, {Kubo}, \& {IKAROS Demonstration Team}}]{YonEtAl:2011}
{Yonetoku}, D., {Murakami}, T., {Gunji}, S., {et~al.} 2011, \apjl, 743, L30

\bibitem[{{Yonetoku} {et~al.}(2012){Yonetoku}, {Murakami}, {Gunji}, {Mihara},
  {Toma}, {Morihara}, {Takahashi}, {Wakashima}, {Yonemochi}, {Sakashita},
  {Toukairin}, {Fujimoto}, \& {Kodama}}]{YonEtAl:2012}
---. 2012, \apjl, 758, L1

\bibitem[{{Yu} {et~al.}(2015){Yu}, {van Eerten}, {Greiner}, {Sari}, {Narayana
  Bhat}, {von Kienlin}, {Paciesas}, \& {Preece}}]{YuEtAl:2015}
{Yu}, H.-F., {van Eerten}, H.~J., {Greiner}, J., {et~al.} 2015, \aap, 583, A129

\end{thebibliography}

\appendix

\section{The locally emitted synchrotron spectrum}
\label{sect:A}

The comoving synchrotron emissivity produced by electrons (and positrons) with a Lorentz factor distribution $\mathrm{d}n_\pm/\mathrm{d}\gamma$
is given by

\be
\jp = \frac{1}{4\pi} \int \frac{\D n_\pm}{\D \gamma} \Pp \D \gamma,
\label{eq:jp_1}
\ee

\noindent where $\Pp$ is the spectral power (erg s$^{-1}$ Hz$^{-1}$) emitted by each electron, and we have approximated the emissivity as isotropic. We use primes on the comoving emissivity and frequencies, to distinguish them from the corresponding unprimed lab frame quantities.

We will make the simplifying assumption that each electron emits only at its own characteristic synchrotron frequency $\nu^\prime = \gamma^2 \nucyc$, where $\nucyc \equiv e B / 2 \pi m_e c$ is the Larmor frequency. This gives the spectral power emitted by a single electron in the form $\Pp = (\gamma/\gamma_0)^2 P_0\, \delta(\nup - \gamma^2 \nucyc)$, where $\delta(...)$ is the delta function and $P_0 \equiv \gamma_0^2 \sigma_\mathrm{T} c B^2/6 \pi$ is the power emitted by an electron of Lorentz factor $\gamma_0$.

The definite relation between the emitted frequency and the electron Lorentz factor simplifies the integral in \Eq~(\ref{eq:jp_1}), which gives

\be
\jp = \frac{1}{8\pi} \frac{\gamma P_0}{\gamma_0^2 \nucyc} \frac{\D n_\pm}{\D \gamma},
\label{eq:jp_2}
\ee

\noindent where $|\D\nup/\D\gamma| = 2 \gamma \nucyc$ was used to integrate out the delta function.

As the outflow propagates a distance $\D r$, a luminosity $\D L_\nu$ (erg s$^{-1}$ Hz$^{-1}$) is added to the synchrotron spectrum by the injected relativistic electrons. The luminosity is related to the flux throught the sphere of radius $r$ by $\D L_\nu = 4 \pi r^2 \D F_\nu$,
and so we have

\be
\D L_\nu = 4 \pi r^2 \int\limits_{2\pi} \mu \D I_\nu \D\Omega,
\ee

\noindent where $\mu$ is the cosine of the angle to the local radial direction, $\D\Omega$ is a solid angle element and the integration is over the outer half-sphere (i.e. $0 < \mu < 1$, radiation propagating outwards). The added specific intensity $\D I_\nu$ is

\be
\D I_\nu = j_\nu \frac{\D r}{\mu},
\ee

\noindent where $j_\nu$ is the lab frame emissivity. One then finds

\be
\frac{\D L_\nu}{\D\ln r} = 4 \pi r^3 \int\limits_{2\pi} j_\nu \D\Omega.
\label{eq:omega_int}
\ee

The bulk of radiation produced by the jet is radially beamed within angles $\delta\theta\sim \Gamma^{-1}$, and this radiation is assumed to have axial symmetry about the radial direction. Integration over the azimuthal angle is then performed by replacing $\D\Omega = 2\pi \D\mu$. The lab frame frequency is a function of $\nup$ and $\mu$; $\nu = D\nup$ where $D \equiv (\Gamma[1-\beta\mu])^{-1}$ is the Doppler boost and $\beta$ is the outflow speed in units of the speed of light. Since $\nup$ is a function of $\gamma$, we may change the variable of integration in \Eq~(\ref{eq:omega_int}) to $\gamma$ for a constant $\nu$,

\be
\D\Omega = 4\pi \frac{\gamma\nucyc}{\Gamma\nu} \D\gamma.
\label{eq:dOmega}
\ee

\noindent The upper limit of integration is $\gamma_0$ and corresponds to the lower limit in $\mu$ (electrons need a larger $\gamma$ to emit at frequency $\nu$ if they emit at smaller $\mu$). The lower limit corresponds to electrons emitting radially, as this is the direction of the largest Doppler boost. For the radial direction we have $\nu = 2\Gamma\nup = 2 \Gamma \gamma_\mathrm{min}^2 \nucyc$, or $\gamma_\mathrm{min} = (\nu/2\Gamma\nucyc)^{1/2} = \gamma_0(\nu/\numax)^{1/2}$, where

\be
\numax \equiv 2\Gamma\gamma_0^2\nucyc
\ee

\noindent is the highest frequency of emission in the lab frame. The transformation of the emissivity is $j_\nu = D^2 \jp$, and we have $\nu = D \nup = D \gamma^2 \nucyc$, so that $j_\nu = \jp (\nu / \gamma^2\nucyc)^2$. Combining equations (\ref{eq:jp_2}) to (\ref{eq:dOmega}) above, we find

\be
\nu\frac{\D L_\nu}{\D\ln r} = 4 \pi r^3 \frac{2\Gamma \gamma_0^2 P_0\nu^2}{\numax^2} \int\limits_{\gamma_0(\nu/\numax)^{1/2}}^{\gamma_0} \frac{1}{\gamma^2} \frac{\D n_\pm}{\D \gamma} \D\gamma.
\label{eq:nudLnudlnr}
\ee

\noindent The unscattered synchrotron spectrum is then

\be
\nu L_\mathrm{\nu, nsc} = \int \nu\frac{\D L_\nu}{\D\ln r} \exp(-\tau) \D\ln r,
\label{eq:nudLnudlnr_nsc}
\ee

\noindent where $\tau$ is the sum of the optical depths of Thomson scattering and synchrotron self-absorption. In order to compute the unscattered synchrotron emission spectrum, we numerically integrate equations (\ref{eq:nudLnudlnr}) and (\ref{eq:nudLnudlnr_nsc}), taking $\mathrm{d}n_\pm/\mathrm{d}\gamma$ and $\Gamma$ (as functions of radius) from the full radiative transfer simulations.

\section{The approximate electron Lorentz factor distribution}
\label{sect:B}

Due to the rapid cooling of the injected electrons (and positrons), the electron Lorentz factor distribution is approximately locally time independent, and can be found by solving the kinetic equation

\be
\frac{\mathrm{d}}{\mathrm{d}\gamma}\left(\dot{\gamma} \frac{\mathrm{d}n_\pm}{\mathrm{d}\gamma}\right) + S(\gamma) = 0,
\ee

\noindent or

\be
\frac{\mathrm{d}n_\pm}{\mathrm{d}\gamma} = -\frac{1}{\dot{\gamma}} \int_\gamma^{\gamma_0} S(\gamma) \mathrm{d}\gamma,
\label{eq:7}
\ee

\noindent where $S(\gamma)$ is a source term that describes the injection of nonthermal, primary electrons, $\dot{\gamma} m_e c^2 = -(P_\mathrm{syn} + P_\mathrm{IC}) = -P_\mathrm{syn} (\eB + \KN\erad) / \eB$ describes the cooling of the electron by both synchrotron emission and scatterings, and $P_\mathrm{syn} = (\gamma/\gamma_0)^2 P_0$. If we assume that all primary electrons are injected at $\gamma = \gamma_0$, then $S = \dot{n}^\mathrm{inj}_\pm \delta(\gamma - \gamma_0)$, where $\dot{n}^\mathrm{inj}_\pm$ is the injection rate of electrons per volume. Performing the integration in \Eq~(\ref{eq:7}), we find

\be
\frac{\mathrm{d}n_\pm}{\mathrm{d}\gamma} = \frac{\eB}{\eB + \KN\erad} \frac{m_e c^2 \gamma_0^2 \dot{n}^\mathrm{inj}_\pm}{\gamma^2 P_0}.
\label{eq:8}
\ee

\noindent The energy injection rate into the plasma can be written as

\be
\dot{Q} = \gamma_0 m_e c^2 \dot{n}^\mathrm{inj}_\pm,
\ee

\noindent and is related to the dissipated luminosity per logarithmic interval in radius, $\D\Ld/\D\ln r$, by

\be
\dot{Q} = \frac{1}{4\pi r^3 \Gamma} \frac{\mathrm{d}\Ld}{\mathrm{d}\ln r}.
\label{eq:Q}
\ee

\noindent Combining equations (\ref{eq:8}) to (\ref{eq:Q}), we find

\be
\frac{\mathrm{d}n_\pm}{\mathrm{d}\gamma} = \frac{\eB}{\eB + \KN\erad} \frac{\gamma_0}{4 \pi r^3 \Gamma \gamma^2 P_0} \frac{\D\Ld}{\D\ln r}.
\label{eq:n_pm}
\ee

\noindent Inserting \Eq~(\ref{eq:n_pm}) into \Eq~(\ref{eq:nudLnudlnr}) and integrating over $\gamma$, we obtain the locally emitted synchrotron spectrum,

\be
\nu \frac{\mathrm{d}L_\nu}{\mathrm{d}\ln r} = \frac{2}{3} \frac{\eB}{\eB + \KN\erad} \frac{\mathrm{d}L_\mathrm{d}}{\mathrm{d}\ln r} \left(\frac{\nu}{\nu_\mathrm{max}}\right)^{1/2} \left[1 - \left(\frac{\nu}{\nu_\mathrm{max}}\right)^{3/2}\right].
\label{eq:dLnudlnr_simple}
\ee

\noindent This spectrum could also have been obtained by simply considering that the locally emitted spectrum should have $L_\nu \propto \nu^{-1/2}$ due to fast cooling electrons, extend up to $\numax \approx \Gamma \nup_\mathrm{syn}(\gamma_0)$ and the total energy emitted in synchrotron emission is a fraction $\eB/(\eB + \KN\erad)$ of the total dissipated nonthermal energy (which is obtained by integration of \Eq~(\ref{eq:dLnudlnr_simple}) over $\nu$).

\section{Optical depth due to synchrotron self-absorption}
\label{sect:C}

The angle-averaged synchrotron self-absorption coefficient is (e.g. \citealt{GhiSve:1991, VurEtAl:2011})

\be
\kappa_\mathrm{\nu^\prime} = -\frac{1}{2 m_e (\nu^\prime)^2} \int \frac{P_\mathrm{\nu^\prime}}{4\pi} \gamma p \frac{\D}{\D p} \left(p^{-2} \frac{\D n_\pm}{\D p}\right) \D p,
\label{eq:kappa}
\ee

\noindent where $p \equiv \beta\gamma$ is the dimensionless electron momentum. For high energy power law electrons, which dominate the synchrotron self-absorption opacity at the frequencies of interest, we have $p \approx \gamma$. As above, we approximate the spectral power from a single electron as $P_\mathrm{\nu^\prime} \approx (\gamma/\gamma_0)^2 P_0 \delta(\nu^\prime - \gamma^2 \nucyc)$, which gives

\be
\kappa_\mathrm{\nu^\prime} \approx -\frac{P_0}{16\pi m_e (\nucyc)^3 \gamma\gamma_0^2} \frac{\D}{\D \gamma} \left(\gamma^{-2} \frac{\D n_\pm}{\D \gamma}\right).
\label{eq:kappa_2}
\ee

\noindent The synchrotron self-absorption optical depth at a given lab frame frequency $\nu \approx \Gamma\gamma^2\nucyc$ is then $\tau_{\nu^\prime} \approx \kappa_\mathrm{\nu^\prime} r / \Gamma$. \Eq~(\ref{eq:kappa_2}) can be evaluated numerically for a given electron Lorentz factor distribution. For estimates, one may use the approximate electron Lorentz factor distribution of \Eq~(\ref{eq:n_pm}). Given the dissipation rate of \Eq~(\ref{eq:dLdlnr}) and the approximate electron Lorentz factor distribution, we find

\be
\tau_\mathrm{\nu^\prime} \approx \frac{\eB}{\eB + \KN\erad} \frac{\eph L}{(4\pi)^2 r^2 (\nu^\prime)^3 \Gamma^2 \gamma_0 m_e} \left(\frac{r}{\Rph}\right)^k.
\label{eq:tau_nu}
\ee

\label{lastpage}

\end{document}